\documentclass{statsoc}\usepackage[]{graphicx}\usepackage[]{color}
\makeatletter
\def\maxwidth{ %
  \ifdim\Gin@nat@width>\linewidth
    \linewidth
  \else
    \Gin@nat@width
  \fi
}
\makeatother

\definecolor{fgcolor}{rgb}{0.345, 0.345, 0.345}
\newcommand{\hlnum}[1]{\textcolor[rgb]{0.686,0.059,0.569}{#1}}%
\newcommand{\hlopt}[1]{\textcolor[rgb]{0,0,0}{#1}}%
\newcommand{\hlstd}[1]{\textcolor[rgb]{0.345,0.345,0.345}{#1}}%

\usepackage{framed}
\makeatletter
\newenvironment{kframe}{%
 \def\at@end@of@kframe{}%
 \ifinner\ifhmode%
  \def\at@end@of@kframe{\end{minipage}}%
  \begin{minipage}{\columnwidth}%
 \fi\fi%
 \def\FrameCommand##1{\hskip\@totalleftmargin \hskip-\fboxsep
 \colorbox{shadecolor}{##1}\hskip-\fboxsep
     \hskip-\linewidth \hskip-\@totalleftmargin \hskip\columnwidth}%
 \MakeFramed {\advance\hsize-\width
   \@totalleftmargin\z@ \linewidth\hsize
   \@setminipage}}%
 {\par\unskip\endMakeFramed%
 \at@end@of@kframe}
\makeatother

\definecolor{shadecolor}{rgb}{.97, .97, .97}
\definecolor{messagecolor}{rgb}{0, 0, 0}
\definecolor{warningcolor}{rgb}{1, 0, 1}
\definecolor{errorcolor}{rgb}{1, 0, 0}
\newenvironment{knitrout}{}{} 

\usepackage{alltt}

\usepackage[a4paper]{geometry}
\usepackage{graphicx}
\usepackage[textwidth=8em,textsize=small]{todonotes}
\usepackage{amsmath}
\usepackage{natbib}
\usepackage{textcomp}

\title[Parsimonious Mixed Models]{Parsimonious Mixed Models}

\author{Douglas Bates}
\address{Statistics, University of Wisconsin-Madison, Madison, USA.}
\email{bates@stat.wisc.edu}
\author{Reinhold Kliegl}
\address{Psychology, University of Potsdam, Potsdam, Germany.}
\author{Shravan Vasishth}
\address{Linguistics, University of Potsdam, Potsdam, Germany,}
\author[Bates, Kliegl, Vasishth, Baayen]{R. Harald Baayen}
\address{Linguistics, University of T\"ubingen, T\"ubingen, Germany\\
Linguistics, University of Alberta, Edmonton, Canada}
\IfFileExists{upquote.sty}{\usepackage{upquote}}{}
\begin{document} 
 \begin{abstract}
  The analysis of experimental data with mixed-effects models requires decisions about the specification of the appropriate random-effects structure. Recently, Barr, Levy, Scheepers, and Tily 2013 recommended  fitting `maximal' models with all possible random effect components included.  Estimation of maximal models, however, may not converge.  We show that failure to converge typically is not due to a suboptimal estimation algorithm, but is a consequence of attempting to fit a model that is too complex to be properly supported by the data, irrespective of whether estimation is based on maximum likelihood or on Bayesian hierarchical modeling with uninformative or weakly informative priors.  Importantly, even under convergence, overparameterization may lead to uninterpretable models.  We provide diagnostic tools for detecting overparameterization and guiding model simplification.
  \end{abstract}
 
 \keywords{linear mixed models, model selection, crossed random effects, model simplicity}

 \section{Introduction}

During the last ten years, there has been a significant change in how psycholinguistic experiments are analyzed when both subjects and items are included as random factors, specifically a change from analyses of variance to linear mixed models (LMMs), with \citet{BaayenDavidsonBates2008} providing a first major introduction.
Although hierarchical linear models have been in existence for decades, their adoption in areas like psychology and linguistics became more widespread in areas like linguistics and psychology after \citet{PB} appeared \citep{vasishthphdbook,VasishthLewisLanguage06,BaayenDavidsonBates2008,OberauerKliegl06,KlieglRisseLaubrock2007,Kliegl2007}. Recently, the use of LMMs has spread to other areas of psychology, such as personality and social psychology \citep{JuddWestfallKenny2012,WestfallKennyJudd2014}. There are a number of reasons for this change. One particularly attractive feature has been that, with LMMs, statistical inference about experimental effects and interactions no longer needs separate analyses of variance, one for subjects and one for items \citep{Clark1973,ForsterDickinson1976}, but can be carried out within a single coherent framework. 

This benefit in coherence comes at some cost. An important part of analyzing experimental data with mixed-effects models is the selection of the proper random-effects structure.  In principle, LMMs not only consider variance between subjects and between items in the mean of the dependent variable (i.e., random intercepts), but also variance between subjects and between items for all main effects and interactions (i.e., random slopes) as well as correlations between intercepts and slopes. Let us illustrate the basic point with the most simple model with a two-level within-subject experimental manipulation and subject as the only random factor, using the formula notation of the `lme4`
package for R described in \citet{BatesMaechlerBolkerWalker2015}:

\begin{knitrout}
\definecolor{shadecolor}{rgb}{0.969, 0.969, 0.969}\color{fgcolor}\begin{kframe}
\begin{alltt}
\hlstd{Y} \hlopt{~} \hlnum{1} \hlopt{+} \hlstd{A} \hlopt{+} \hlstd{(}\hlnum{1}\hlopt{|}\hlstd{Subject)}
\end{alltt}
\end{kframe}
\end{knitrout}

This model may support the fixed effect of the within-subject factor A. However, if the effect of A (i.e., the difference between the two experimental conditions) differs reliably between subjects, uncertainty about A may be so substantial that the main effect of A is no longer significant in a model allowing for random slopes for A:

\begin{knitrout}
\definecolor{shadecolor}{rgb}{0.969, 0.969, 0.969}\color{fgcolor}\begin{kframe}
\begin{alltt}
\hlstd{Y} \hlopt{~} \hlnum{1} \hlopt{+} \hlstd{A} \hlopt{+} \hlstd{(}\hlnum{1}\hlopt{+}\hlstd{A}\hlopt{|}\hlstd{Subject)}
\end{alltt}
\end{kframe}
\end{knitrout}

For the assessment of the significance of the experimental manipulation, it is therefore essential to examine its fixed effect in the presence of the corresponding random slopes (see, e.g., \citet{PB},\citet{Baayen2008}). 

\section{Parameter estimation in mixed models} \label{pemm}

Following the publication of \citet{BarrLevyScheepersTily13} containing the advice in its title to ``keep it maximal'' when formulating an LMM for confirmatory analysis, the frequency of reports and queries related to failure of a model fit to converge has increased, for example, in the discussion list for the \texttt{lme4} package for R \citep{R-lme4}.
Although LMMs and generalized linear mixed-effects models (GLMMs) are versatile tools for modeling the variability in observed responses and attributing parts of this variability to different sources, like any statistical modeling technique they have their limitations.  Knowledge of these limitations is important in ensuring appropriate usage. 

As the complexity of the model increases, so does the difficulty of the optimization problem. For LMMs, aside from intercept and fixed effects, the parameters being estimated represent variances and covariances, which are typically much more difficult to estimate than regression coefficients. The parameters in GLMMs are even more complicated.

When there is more than one experimental factor, say in a factorial design, the number of parameters to be estimated explodes.  For example, a maximal model with three experimental within-subject and within-items factors with two levels each in a full factorial design incorporating the seven main effects and interactions plus the intercept with random effects for subject and item, would require estimation of eight fixed-effects coefficients
and $73$ variance-covariance parameters.  The online supplement to
\citet{BarrLevyScheepersTily13} fits such a model to data from an
experiment described in \citet{KronmullerBarr2007}. We also present a reanalysis of this experiment below.
  
To anticipate the main result, it is simply not realistic to try to fit this number of highly abstract parameters given the number of subjects and items in this experiment.  Almost unfortunately, the software does indeed converge to parameter estimates but these estimates correspond to degenerate or singular covariance matrices, in which some linear combinations of the random effects are estimated to having no variability. This corresponds to estimates of zero random-effects variance in a model with random-intercepts
only or a correlation of $\pm 1$ in a model with correlated random intercepts and slopes.  However, already a three-by-three correlation matrix will not usually show boundary values like these, even when it is singular. In summary, the parameters representing variances and covariances are constrained in
complicated ways.  In overparameterized models, convergence can occur on the boundary, corresponding to models with singular variance-covariance matrices for random effects. 
This can have serious, adverse consequences for inference; for example,  due to an overparameterization of the maximal LMM, \citet{KlieglWeiDambacherYanZhou2011} wrongly interpreted an LMM correlation parameter as providing much more evidence than the corresponding within-subject correlation for the correlation of two experimental effects.

In a linear mixed model incorporating vector-valued random effects, say by-subject random effects for intercept and for slope, the variance component parameters determine a variance-covariance matrix for these random effects.  As described in \citet{BatesMaechlerBolkerWalker2015}, the parameters used
in fitting the model are the entries in the Cholesky factor ($\Lambda$) of the relative variance-covariance matrix of the unconditional distribution of the random effects.  The parameter vector ($\theta$) for this model are the values on and below the diagonal of a lower triangular Cholesky factor.  The $\theta$
vector elements fill the lower triangular matrix in column major order. The relative covariance matrix for the random effects is $\Lambda\Lambda^\prime$. To reproduce the covariance matrix, $\Lambda\Lambda^\prime$ must be scaled by $s^2$.

When one or more columns of the Cholesky factor $\Lambda$ are zero vectors, $\Lambda$ is rank-deficient: The linear subspace formed by all possible linear combinations of the columns is of reduced dimensionality compared to the dimensionality of $\Lambda$.  The random-effects vectors that can be generated from this fitted model must lie in this lower-dimensional subspace. That is, there will be no variability in one or more directions of the space of random
effects.

The \texttt{RePsychLing} package provides a new function, \texttt{rePCA} (which may
become part of a future release of `lme4`) that enables the analyst to probe models fitted with \texttt{lmer} for rank deficiency.  The \texttt{rePCA} (random-effects Principal Components Analysis) function takes an object of class \texttt{lmerMod} (i.e. a model fit by \texttt{lmer}) and produces a list of principal component (\texttt{prcomp}) objects, one for each grouping factor. These principal component objects can be summarized and visualized (by means of scree plots), exactly as any other principal component object generated by the
\texttt{prcomp} function of R.  

Principal components analysis of the estimated covariance matrices for the random effects in a linear mixed model allows for simple assessment of the dimensionality of the random effects distribution. As illustrated below and in the vignettes in the  \texttt{RePsychLing} package, the maximal model in many analyses of data from Psychology and Linguistics experiments, is almost always
shown by this analysis to be degenerate.

\section{Iterative reduction of model complexity}

In this section, we assume that the researcher has hypotheses about main effects and (some) interactions, but that he/she has no specific expectations about variance components or correlation parameters. In other words, we assume that the experimental hypotheses relate to the fixed effects, not the random-effects structure. In hypothesis testing, usually  the primary reason
for dealing with the random-effects structure is to obtain as powerful tests as justified of the fixed effects. Therefore, it is reasonable to remove variance components/correlation parameters from the model if they are not supported by the data. If there are specific expectations about, say, a correlation parameter, it makes sense to include it in the model (as well as the related variance components). We also assume the standard situation whereby potential numeric within-subject or within-item covariates are not under consideration.

In a factorial experiment, the maximum number of variance-covariance parameters to be estimated for each random factor is  

\begin{equation*}
\frac{(\hbox{product of within-factor levels}) \times (\hbox{product of within-factor levels} + 1)}{2}.
\end{equation*}

For example, a $2\times 2$ within-factor design will have $(2\times 2) \times ((2\times 2)+1)/2=10$ parameters in the variance-covariance matrix for each random effect (commonly, subject and item), and a $2\times 2\times 2$
within-factor design will have $36$ parameters.  Additional model parameters are required for the fixed effects intercept, main effects, interactions, and for the residual variance. It is reasonable to start by attempting to fit a maximal LMM. If this model converges within reasonable time, several steps can be taken to check the possibility of an iterative reduction of model complexity in order to arrive at a parsimonious LMM. 

First, it is worth checking whether we can reduce the dimensionality of the variance-covariance matrices assumed in a
maximal LMM.  The number of principal components that cumulatively
account for 100\% of the variance is a reasonably stringent criterion for settling on the reduced dimensionality. This can be achieved by performing PCA using the \texttt{rePCA()} function on the fitted maximal LMM (see section~\ref{pemm} above).

Second, after we have determined the number of dimensions supported by the data, we can eliminate variance components from the LMM, following the standard statistical principle with respect to interactions and main effects: variance components of higher-order interactions should generally be taken out of the model before
lower-order terms nested under them. Frequently, in the end, this leads also to the elimination of variance components of main effects. The reduced model may be submitted again to a PCA to check the dimensionality of the random-effects structure.

Third, we can check whether forcing to zero the correlation parameters of the reduced LMM significantly decreases the goodness of fit according to a likelihood ratio test (LRT), possibly also taking into account changes in AIC and BIC. Obviously, if the goodness of fit does not change from the reduced model to the zero-correlation-parameter (ZCP) model, we do not have
reliable evidence that the correlation parameters are different from zero. Importantly, this does not mean that the correlations are zero, only that we do not have enough evidence for them being different from zero for the current data; absence of evidence is not evidence of absence. Also, note that the estimated value of the correlation parameters depends on the choice of contrasts for the experimental factors. For example, treatment contrasts and
sum contrasts may lead to models with a very different random-effect structure (see http://www.rpubs.com/Reinhold/22193; this is also available as a vignette in the \texttt{RePsychLing} package).  It is also conceivable that correlation parameters are zero for only one of several random factors. The new $`||'$ syntax of `lmer()' is very convenient for specifying such ZCP LMMs.
However, as illustrated in the same vignette, there are a few constraints relating to general R-formula syntax one needs to know about. In general, the $`||`$ syntax works (currently) as expected only for LMMs after converting factor-based to vector-valued random-effects structures. 

Fourth, we may also want to check whether all the variance components of an identified  model are necessary. Taking out one term at a time and checking again whether there is a significant drop in goodness of fit, allows us to identify variance components that are not supported by the data. Again, removing such terms does not mean that the variance is zero, only that we have no evidence of it being significantly different from zero.  

Fifth, after removing non-significant variance components, we may want to recheck whether the goodness of fit of the iteratively reduced model increases if it is extended with correlation parameters. A reliable variance parameter (i.e., a variance parameter contributing significantly to the goodness of fit according to a likelihood-ratio test) is a necessary condition for estimating correlation parameters associated with this variance component. In other words, we expect to find statistically significant correlation parameters only if the related variance components are statistically significant by themselves.

Finally, as a special case, the iterative reduction of model complexity described above assumed that there was a solution for the maximal model. With complex experiments, however, it may happen that the maximal model does not converge to a solution (e.g., there are warnings that no solution was found) or that the solution is obviously degenerate (i.e., variance components or correlation parameters are estimated at their boundaries of $0$ or $\pm 1$, respectively). In this case, a first step could be to check the dimensionality of the zero-correlation parameter model with a PCA. Obviously, with complex experiments the switch from maximal to zero-correlation parameter model will yield the largest simplification. As already mentioned, whenever one switches from maximal to zero-correlation parameter models, it is very important to have a clear understanding of the contrast specifications chosen for the experimental factors.

In the following two sections, we describe iterative reductions of LMM complexity for two experiments using the above checks. The data for the first example are from an experiment on pragmatic comprehension of instructions (\citealp{KronmullerBarr2007}, 
Exp.\ 2; reanalyzed with an LMM in \citealp{BarrLevyScheepersTily13}).  The second data set is from a visual-attention experiment \citep{KlieglKuschelaLaubrock2014}, following up a different report with the overparameterized model \citep{KlieglWeiDambacherYanZhou2011}. Detailed reports (including R code) for these analyses are described in vignettes, along with four additional examples. We emphasize that we do not claim that our illustrations are the only way to carry out these analyses, but the strategy outlined above has yielded satisfactory results with all data sets we have analyzed so far. There is no cook-book substitute for theoretical considerations and developing statistical understanding. Each data-set deserves the exercise of judgement on part of the researcher.

\subsection{Reanalysis of Kronm\"uller and Barr (2007)}

Here we apply the iterative reduction of LMM complexity to
truncated response times of a $2\times 2\times 2$ factorial psycholinguistic experiment \citep{KronmullerBarr2007}. This is their Exp.\ 2, reanalyzed with an LMM in \citet{BarrLevyScheepersTily13}. The data are from $56$ subjects who responded to $32$ items. Specifically, subjects had to select one of several objects presented on a monitor with a cursor. The manipulations involved (1) auditory instructions that maintained or broke a precedent of reference for the objects established over prior trials, (2) with the instruction being presented by the speaker who established the precedent (i.e., an old speaker) or a new speaker, and (3) whether the task had to be performed without or with a cognitive load consisting of six random digits. All factors were varied within subjects and within items. There were main effects of Load (L), Speaker (S), and
Precedent (P); none of the interactions were significant. Although standard errors of fixed-effect coefficents varied slightly across models, our reanalyses afforded the same statistical inference about the experimental manipulations as the original article, irrespective of LMM specification (see Figure~\ref{fig:fixefkb} a comparison of fixed effects of maximal and parsimonious LMMs). The purpose of the analysis is to illustrate an assessment of model complexity as far as variance components and correlation parameters are concerned, neither of which were the focus of the original publication. 

\subsubsection{Maximal linear mixed model}

A full factorial model in the fixed-effects can be described by the formula 

\begin{knitrout}
\definecolor{shadecolor}{rgb}{0.969, 0.969, 0.969}\color{fgcolor}\begin{kframe}
\begin{alltt}
\hlopt{~}\hlnum{1}\hlopt{+}\hlstd{S}\hlopt{+}\hlstd{P}\hlopt{+}\hlstd{C}\hlopt{+}\hlstd{SP}\hlopt{+}\hlstd{SC}\hlopt{+}\hlstd{PC}\hlopt{+}\hlstd{SPC}
\end{alltt}
\end{kframe}
\end{knitrout}

\begin{figure}[!htbp]
    \centering
\begin{knitrout}
\definecolor{shadecolor}{rgb}{0.969, 0.969, 0.969}\color{fgcolor}
\includegraphics[width=\maxwidth]{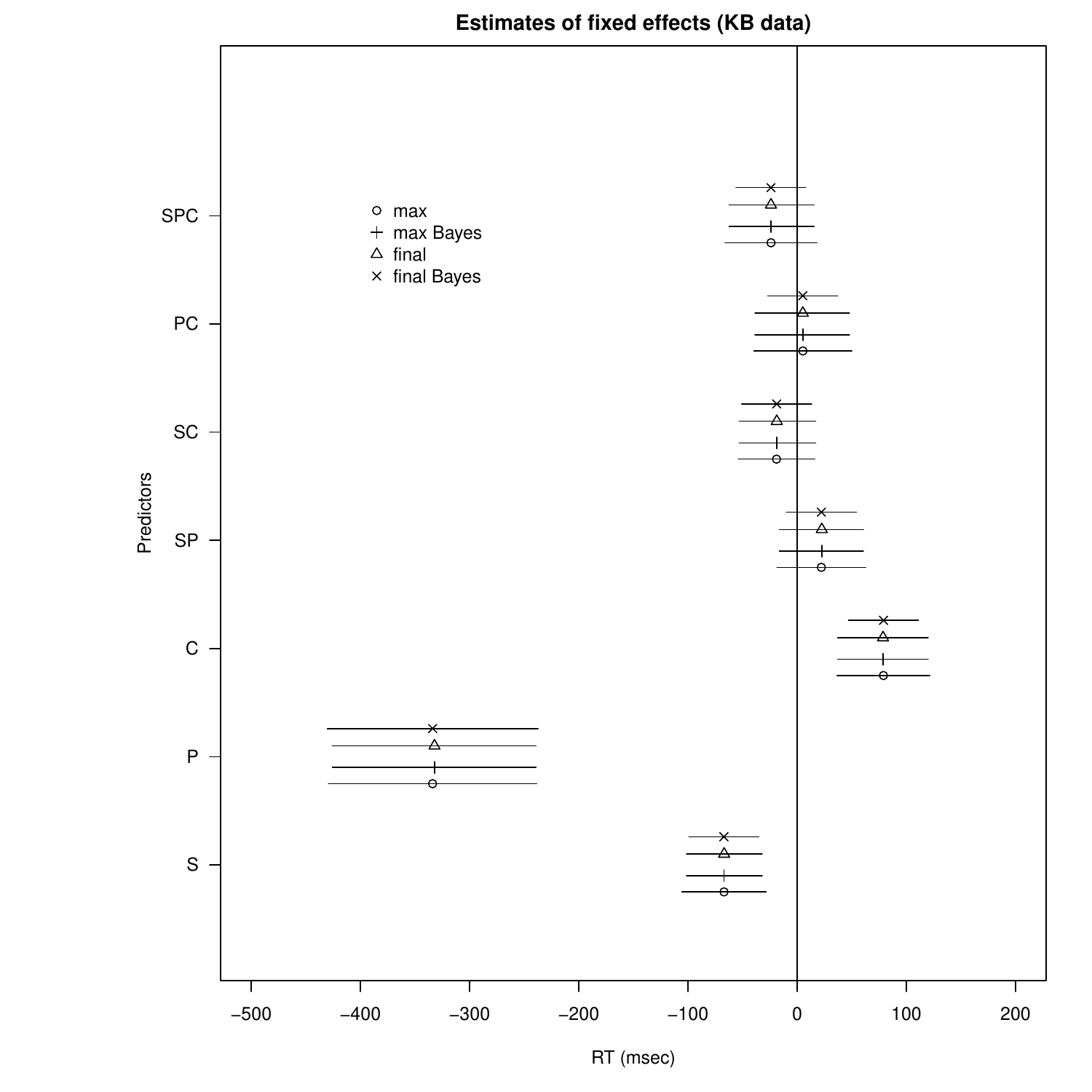} 

\end{knitrout}
\caption{\label{fig:fixefkb} The estimates and 95\% confidence intervals for the fixed effects in the maximal and final models of the Kronm\"uller and Barr 2007 data. Also shown are estimates and 95\% credible intervals from a maximal Bayesian hierarchical linear model.}
\end{figure}

\begin{figure}[!htbp]
    \centering
\begin{knitrout}
\definecolor{shadecolor}{rgb}{0.969, 0.969, 0.969}\color{fgcolor}
\includegraphics[width=\maxwidth]{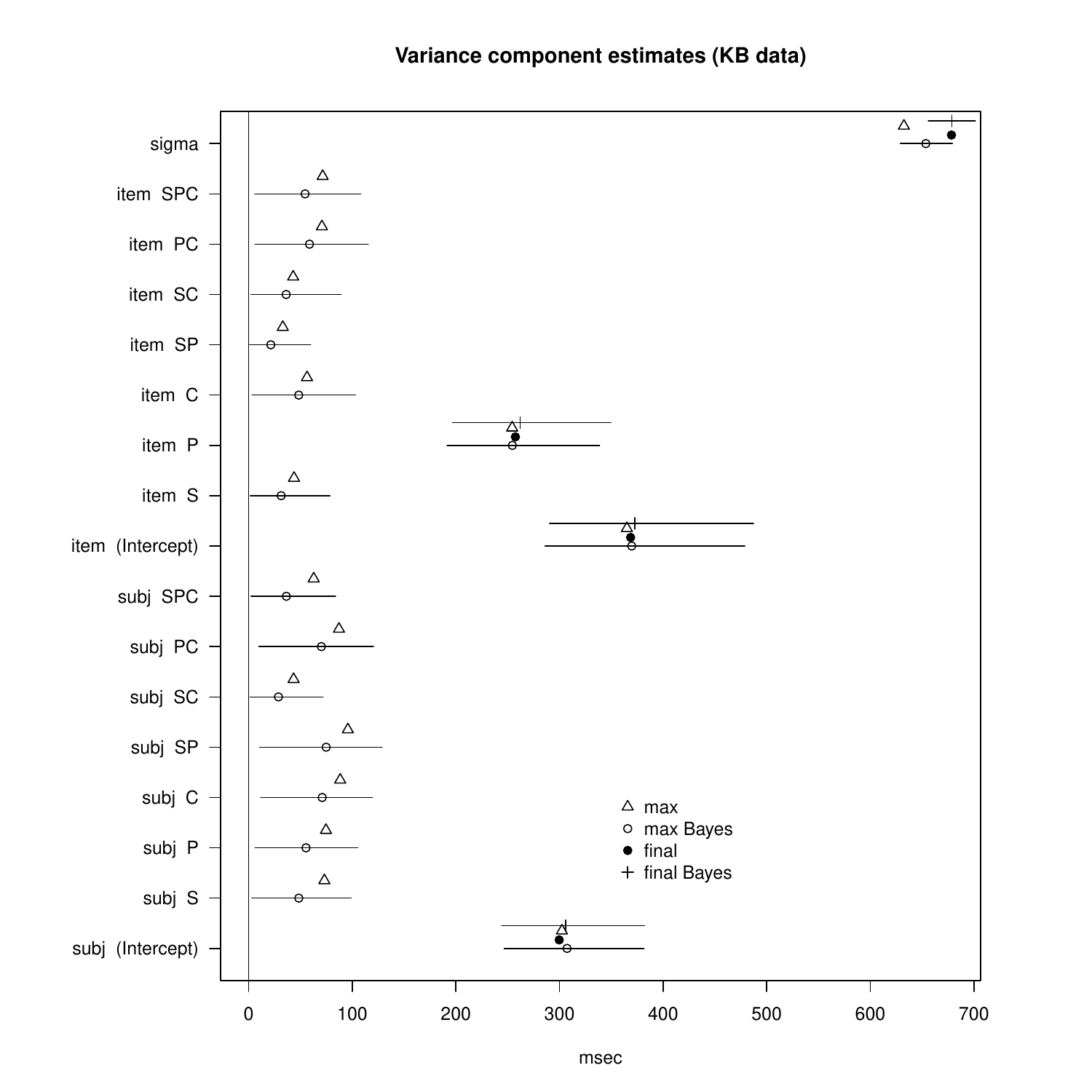} 

\end{knitrout}
\caption{\label{fig:varcompkb} The estimates of the variance components (standard deviations) from the maximal and final models of the KB data. Also shown are the estimates and 95\% credible intervals of the maximal Bayesian hierarchical model.}
\end{figure}

\begin{figure}[!htbp]
    \centering
\begin{knitrout}
\definecolor{shadecolor}{rgb}{0.969, 0.969, 0.969}\color{fgcolor}
\includegraphics[width=\maxwidth]{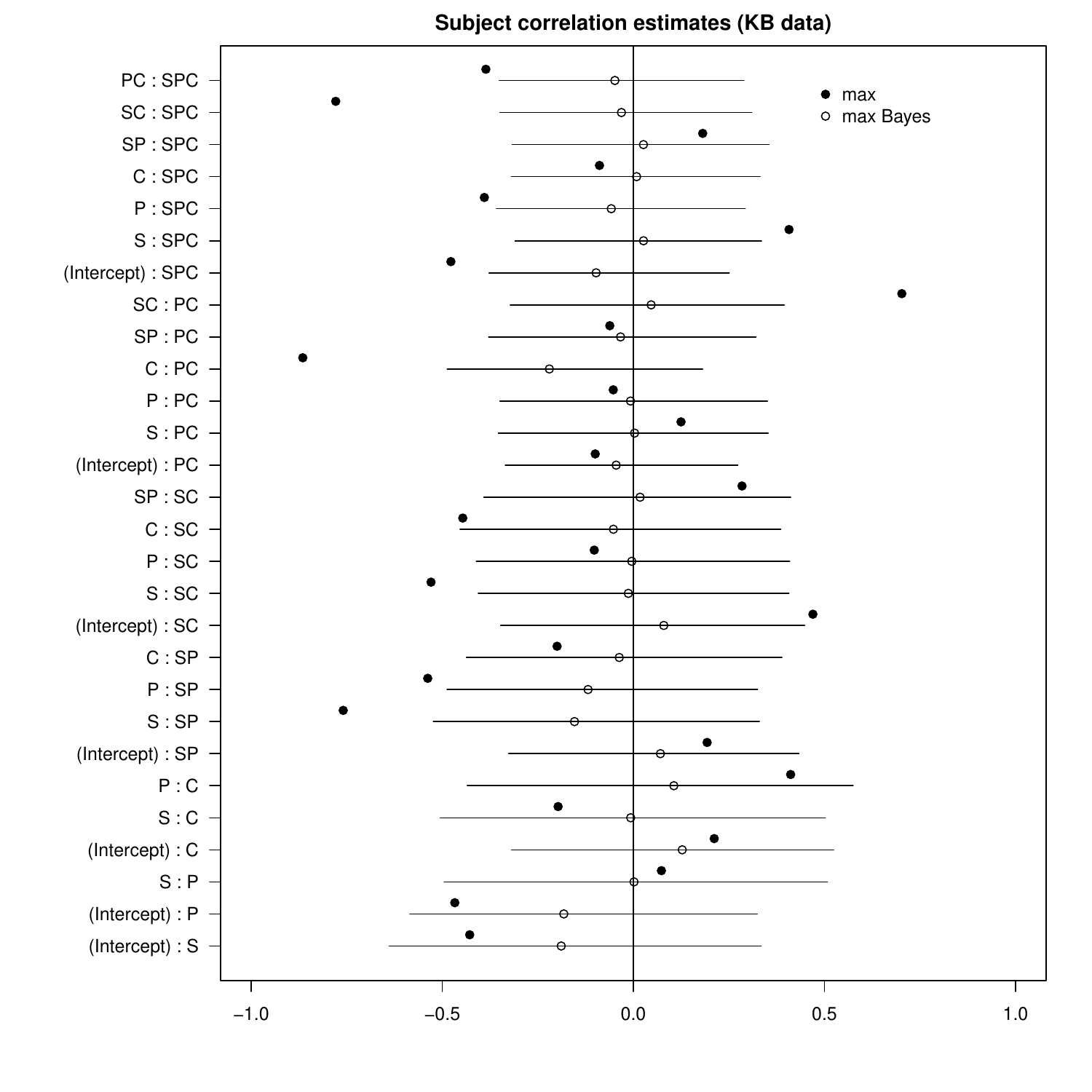} 

\end{knitrout}
\caption{\label{fig:subjcorkb} The estimates of the subject random effect correlations in the lme4 and Bayesian maximal models of the Kronm\"uller and Barr 2007 data.  The Bayesian estimates also have 95\% credible intervals.}
\end{figure}

\begin{figure}[!htbp]
    \centering
\begin{knitrout}
\definecolor{shadecolor}{rgb}{0.969, 0.969, 0.969}\color{fgcolor}
\includegraphics[width=\maxwidth]{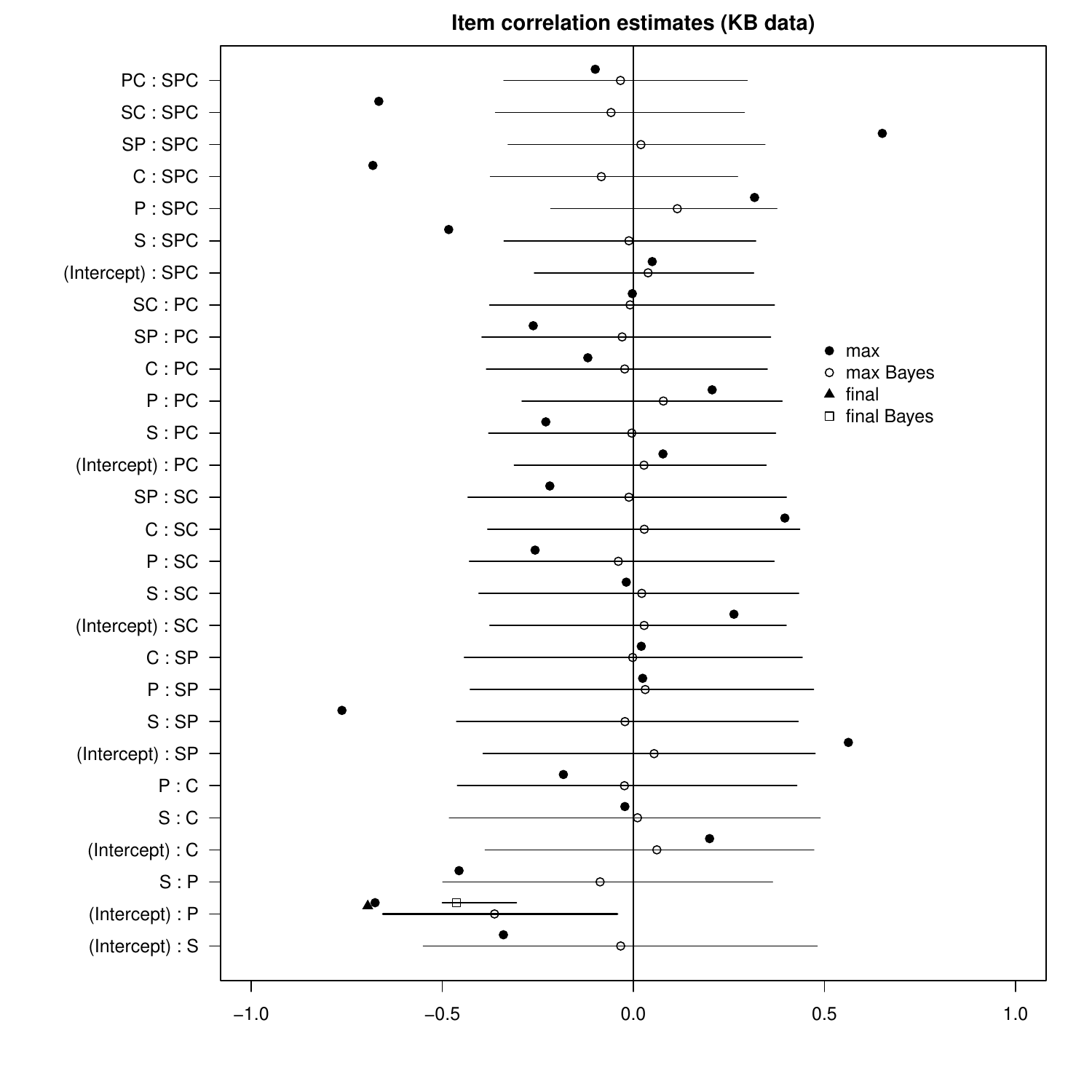} 

\end{knitrout}
\caption{\label{fig:itemcorkb} The estimates of the item random effect correlations in the lme4 and Bayesian maximal models of the Kronm\"uller and Barr 2007 data.  The Bayesian estimates also have 95\% credible intervals.}
\end{figure}

\citet{BarrLevyScheepersTily13} analyzed Kronm\"uller and Barr (2007, Exp.\ 2) with the maximal model for this design comprising $16$ variance components (eight each for the random factors \texttt{SubjID} and \texttt{ItemID}, respectively).  The model took $39,004$ iterations to converge, but produces what look like reasonable parameter estimates (i.e., no variance components with estimates close to zero; no correlation parameters with values close to $\pm 1$). The slow convergence is due to the total of $2 \times 36 = 72$ parameters in the optimization of the random-effects part (ignoring the eight fixed-effect parameters and the residual variance, which do not contribute much to the computational load). Figures  \ref{fig:fixefkb} and \ref{fig:varcompkb} display the fixed effects and variance components of the maximal model, along with other estimates, discussed below. The correlations of subject and item random effects are shown in Figures \ref{fig:subjcorkb} and \ref{fig:itemcorkb}.

Considering that there are only $56$ subjects and $32$ items, it is quite optimistic to expect to estimate $36$ covariance parameters for \texttt{SubjID} and another $36$ for \texttt{ItemID}. A principal components analysis of the variance-covariance matrices for subject and item random effects returns eight principal components, along with the cumulative proportions of variance explained (see Table \ref{tab:kbPCA}).  For subject random effects, four dimensions are sufficient to account for 100\% of the variance explained; and for items, five dimensions suffice. 

\begin{table}
\caption{\label{tab:kbPCA} The cumulative proportion of variance explained for the subject and item random effects in the maximal model for the Kronm\"uller and Barr 2007 data. Principal components analysis was used to compute the cumulative proportion of variance explained.} 
\centering
\begin{tabular}{rr|rrrrrrrr}
& & 1 & 2 & 3 & 4 & 5 & 6 & 7 & 8 \\ 
\hline
subject &  cum. prop. & 0.73 & 0.85 & 0.94 & 1.00 & 1.00 & 1.00 & 1.00 & 1.00 \\ 
item &  cum. prop. & 0.79 & 0.94 & 0.97 & 0.99 & 1.00 & 1.00 & 1.00 & 1.00 \\ 
\end{tabular}
\end{table}

Thus, the maximal model is clearly too complex. In the following paragraphs, we illustrate our iterative method that reduces model complexity to arrive at an optimal LMM for this experiment. We will not report the intermediate results here, but they are available in the vignettes, along with the R code. We qualify this procedure at the outset: We do not claim that this is the only way to proceed, but the strategy has consistently yielded satisfactory results for all data sets we have examined so far. 

\subsubsection{Zero-correlation-parameter linear mixed model}

As a first step toward model reduction, we start with a model including all $16$ variance components, but no correlation parameters. The PCA of this model shows that 12 out of 16 dimensions suffice for capturing 100\% of the variance explained. This suggests that the model is still too complex. 

\subsubsection{Dropping variance components to achieve model identification}

A second step toward model reduction could be to remove variance components to achieve model identification. Starting with the smallest variance component (or a set of them) this step can be repeated until the PCA no longer suggests overidentification. For the present case, variance components 5 and 7 for \texttt{SubjID} and 1 and 4 for \texttt{ItemID} are estimated with zero or close to zero values. We refit the LMM without these variance components. The PCA for this LMM estimates $12$ non-zero variance components.

\subsubsection{Dropping non-significant variance components}

In the third step, we attempt to simplify the random-effects structure of the identified LMM with likelihood ratio tests. For example, the two smallest variance components account for less than 1\% of the variance. We iteratively remove variance components, starting with dropping the highest-order interaction term SPC. Moving on to tests of two-factor interactions, we end up with an LMM comprising only varying intercepts for subject and items and the item-related variance component for P. Looking back to the maximally identified LMM, we see that these are exactly the three variance components with clearly larger standard-deviation estimates ($>249$) compared to the other standard-deviation estimates ($>64$). There is no significant loss of goodness of fit when we remove nine variance components identified this way; $\chi_{9}^{2} = 11.1$, $p = .27$. However, removal of any of three remaining variance components significantly reduces the goodness of fit. 

\subsubsection{Extending the reduced {\sc lmm} with a correlation parameter}

Inclusion of the correlation parameter between item-related intercept and the precedence effect (P) for this model significantly improves the goodness of fit with the correlation parameter estimated at $-0.69$; $\chi_{1}^{2} = 16.3$, $p < .01$. Thus, there is evidence for reliable differences between items in the precedence effect.  The variance components and correlation parameter for this final LMM are displayed in Figure~\ref{fig:varcompkb}.\footnote{Incidentally, although we consider it questionable to compare non-identified and identified models with an LRT, we want to mention that there is no significant difference in goodness of fit between the final LMM and the maximal model we started with. The final number of  principal components suggested by the final model is actually smaller than suggested by the initial PCA of maximal model.} 

\subsubsection{Summary}

In our opinion, the final model we settled on is the \textit{optimal} LMM for the data of this experiment. To summarize our general strategy: (1) we started with a maximal model; (2) then, we fit a zero-correlation model; (3) next, we removed variance components  until the likelihood ratio test showed no further improvement; and (4) finally, we added correlation parameters for the remaining variance components. Principal components analysis
was used throughout to check the dimensionality for the respective intermediate models. This approach worked quite well in the present case. Indeed, we also reanalyzed three additional experiments reported in the supplement of  \citet{BarrLevyScheepersTily13}. As documented in the \texttt{RePsychLing} package accompanying the present article, in each case, the maximal LMM was too complex for the information provided by the experimental data. In each case, the data supported only a very sparse random-effects structure beyond varying intercepts for subjects and items. Fortunately and
interestingly, none of the analyses impacted the statistical inference about fixed effects in these experiments. Obviously, this cannot be ruled out in general. If authors adhere to a strict criterion for significance, such as  $p < .05$ suitably adjusted for multiple comparisons, there is always a chance that a
t-value will fall above or below the criterion across different versions of an LMM.

\subsection{An alternative analysis of Kronm\"uller and Barr (2007) using a Bayesian LMM}

We also show that similar conclusions can be reached if we fit a Bayesian linear mixed model \citep{Gelman2014} instead of the frequentist model discussed above using `lme4`. Presenting the Bayesian estimates corresponding to the maximal linear mixed model presented above provides an independent validation of the conclusion that a simpler model has better motivation. 

We fit a linear mixed model to the  Kronm\"uller and Barr data using \texttt{rstan} 
\citep{rstanpackage}. In a Bayesian linear mixed model, all parameters have a prior distribution defined over them; this is in contrast to the frequentist approach, where each parameter is assumed to have a fixed but unknown value. Defining a prior distribution over each parameter expresses the researcher's existing knowledge about possible values that the parameter can take, before any new data is considered. For example, in the Bayesian formulation, an \texttt{lme4} style model specification such as 

\begin{knitrout}
\definecolor{shadecolor}{rgb}{0.969, 0.969, 0.969}\color{fgcolor}\begin{kframe}
\begin{alltt}
\hlstd{Y} \hlopt{~} \hlnum{1} \hlopt{+} \hlstd{A} \hlopt{+} \hlstd{(}\hlnum{1}\hlopt{|}\hlstd{Subject)}
\end{alltt}
\end{kframe}
\end{knitrout}

\noindent
we could specify that our prior belief about the parameter, call it $\beta_1$, expressing an effect of A is that it has a normal distribution with mean $0$ and some large variance $\sigma^2$. We can write this as

\begin{equation}
  \beta_1 \sim \mbox{Normal}(0,\sigma^2).
\end{equation}

\noindent
Such a prior expresses the belief that, in the absence of any new data, the mean is assumed to be zero, but the large variance expresses uncertainty about this belief. Using computational methods available in \texttt{rstan}, this prior specification can be combined with the data to derive a posterior distribution for each parameter, including the random effects variance components and the correlations between variance components. The posterior distribution of each parameter is effectively a compromise between the prior and the data, and expresses our revised belief about the parameter's distribution after the data are taken into account. If there is strong evidence from the data that the mean of its distribution is different from zero, the mean of the posterior distribution will reflect this. If there is only weak or no evidence from the data---either due to there being too little data or because the mean from the data is near zero---that the parameter has a mean different from zero, then the prior mean of zero will dominate in determining the parameter's posterior distribution. 

The end-product of a Bayesian linear mixed model is always a posterior distribution for each parameter in the model.  Thus, we can plot the 95\% credible interval for each parameter; this interval tells us the range over which we can be 95\% certain that the true value of the parameter lies, given the data. Contrast this with the 95\% confidence interval (CI), which represents one of hypothethically computed CIs over repeated experiments, where 95\% of those hypothetical CIs would contain the true value of the parameter. Note, however, that for relatively large data-sets such as the present one, the credible interval and confidence interval for the fixed effects will generally be identical (see Figure~\ref{fig:fixefkb}).  

We fit a linear mixed model to the Kronm\"uller and Barr data with normal priors on the fixed effects parameters, and a so-called \texttt{lkj} prior on the correlation matrices of the subject and item random effects \citep{stan-manual2015}. The \texttt{lkj} prior assumes that the correlations are zero (with some uncertainty associated with this belief); if there is evidence in the data for a non-zero correlation, the posterior distribution of the correlation parameter will be shifted away from zero.  For the standard deviations, we defined a uniform prior with a bound 0. This prior expresses that we have no strong beliefs about the standard deviation, but we know that it cannot be less than 0 (our analysis does not depend on using this prior). For a detailed specification of the model, see the \texttt{RePsychLing} package.

The posterior distributions of all parameters for the \texttt{maximal} model, along with their 95\% credible intervals, are shown in Figures \ref{fig:fixefkb}, \ref{fig:varcompkb},  \ref{fig:subjcorkb}, and \ref{fig:itemcorkb}. The Bayesian analysis shows two important things. First, the estimates of the fixed effects in the \texttt{lme4} model and the Bayesian model are nearly identical. This shows that the ``maximal'' LMM fit using \texttt{lme4} is essentially equivalent to fitting a Bayesian LMM with regularizing priors of the sort described above.  Second, the relevant variance component parameters that were identified above using principal components analysis (PCA) and likelihood ratio tests  (LRTs) are exactly the parameters that clearly dominate in the Bayesian analysis; see Figures \ref{fig:varcompkb}, \ref{fig:subjcorkb}, and \ref{fig:itemcorkb}. Specifically, any variance component excluded in the \texttt{lme4}-based analysis using PCA and LRTs has, in the Bayesian analysis, a posterior credible interval that includes zero; and any correlation parameter excluded in the \texttt{lme4}-based analysis has, in the Bayesian model, a credible interval that spans zero. In other words, when we approach the analysis from the perspective of Bayesian modeling, we also find that there is no evidence in the data that the relevant parameters have values that are different from zero. These parameters should be excluded from the model on grounds of parsimony. For comparison, we also present Bayesian estimates of the final model with a reduced number of variance components (Figures \ref{fig:fixefkb}-\ref{fig:itemcorkb}).

\subsection{Reanalysis of Kliegl et al.\ (2015)}

As a second demonstration that linear mixed models with a maximal random-effect structure may be asking too much, we re-analyze data from a visual-attention experiment \citep{KlieglKuschelaLaubrock2014}, following up a published experiment \citep{KlieglWeiDambacherYanZhou2011} with an (unfortunately) overidentified LMM, as shown in the vignette for the KWDYZ data in the \texttt{RePsychLing} package. The experiment shows that validly cued targets on a monitor are detected faster than invalidly cued ones (i.e., spatial cueing effect; \citet{Posner1980}) and that targets presented at the opposite end of a rectangle at which the cue had occurred are detected faster than targets presented at a different rectangle but with the same physical distance (object-based effect; \citet{EglyDriverRafal1994}). Different from earlier research, the two rectangles were not only presented in cardinal orientation
(i.e., in horizontal or vertical orientation), but also diagonally
(45\textdegree\ left or 45\textdegree\ right). This manipulation afforded a follow-up of a hypothesis that attention can be shifted faster diagonally than vertically or horizontally across the screen
\citep{KlieglWeiDambacherYanZhou2011,ZhouChuLiZhan2006}. Finally,
data are from two groups of subjects, one group had to detect small targets and the other large targets. The experiment is a follow-up to \citet{KlieglWeiDambacherYanZhou2011} who used only small targets and only cardinal orientations for rectangles. For the interpretation of the fixed effects, we refer to \citet{KlieglKuschelaLaubrock2014}. Again, the different model specifications reported in this section were of no consequence for the significance or interpretation of fixed effects, but they led to inappropriate conclusions about the correlations between variance components. We focus here on
exploring the random-effect structure for these data. 

\subsubsection{Maximal linear mixed model}

\begin{figure}
\centering
\begin{knitrout}
\definecolor{shadecolor}{rgb}{0.969, 0.969, 0.969}\color{fgcolor}
\includegraphics[width=\maxwidth]{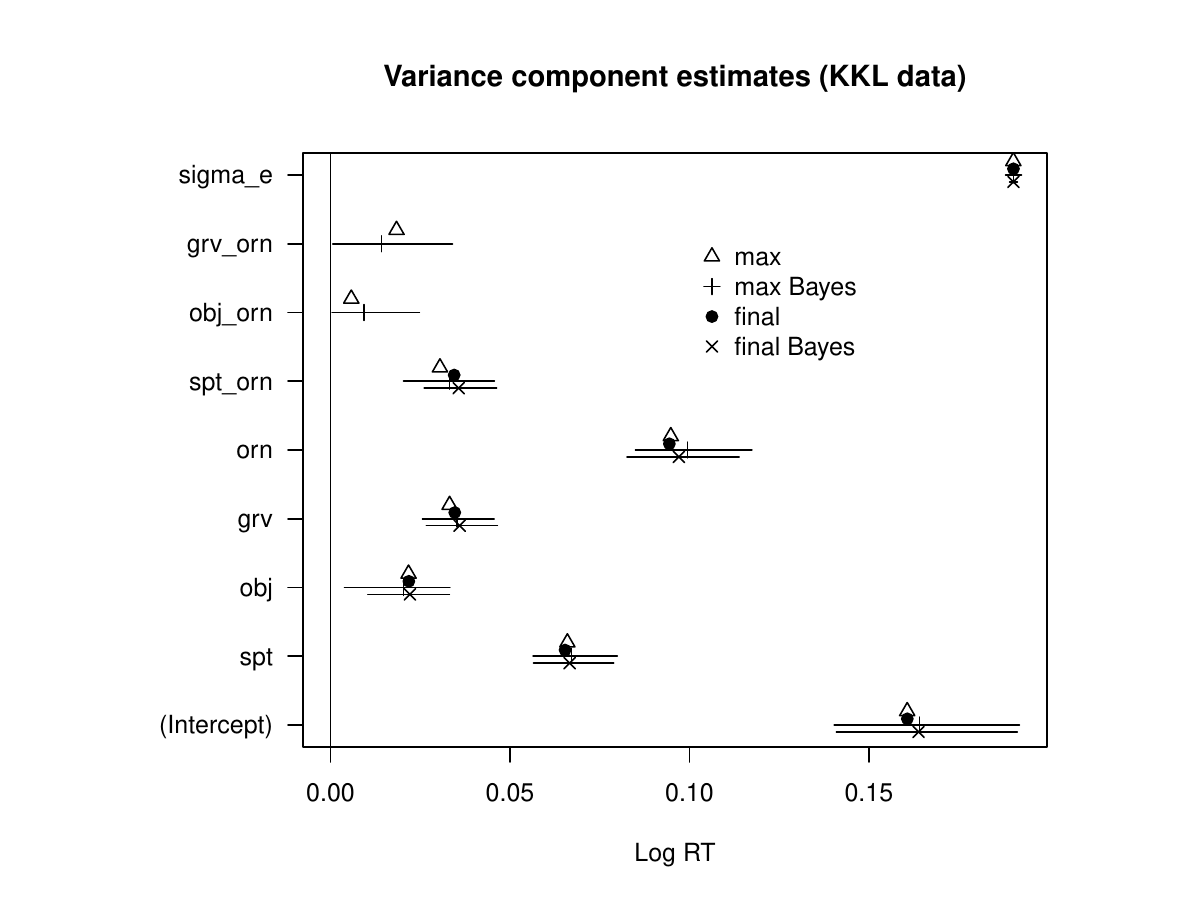} 

\includegraphics[width=\maxwidth]{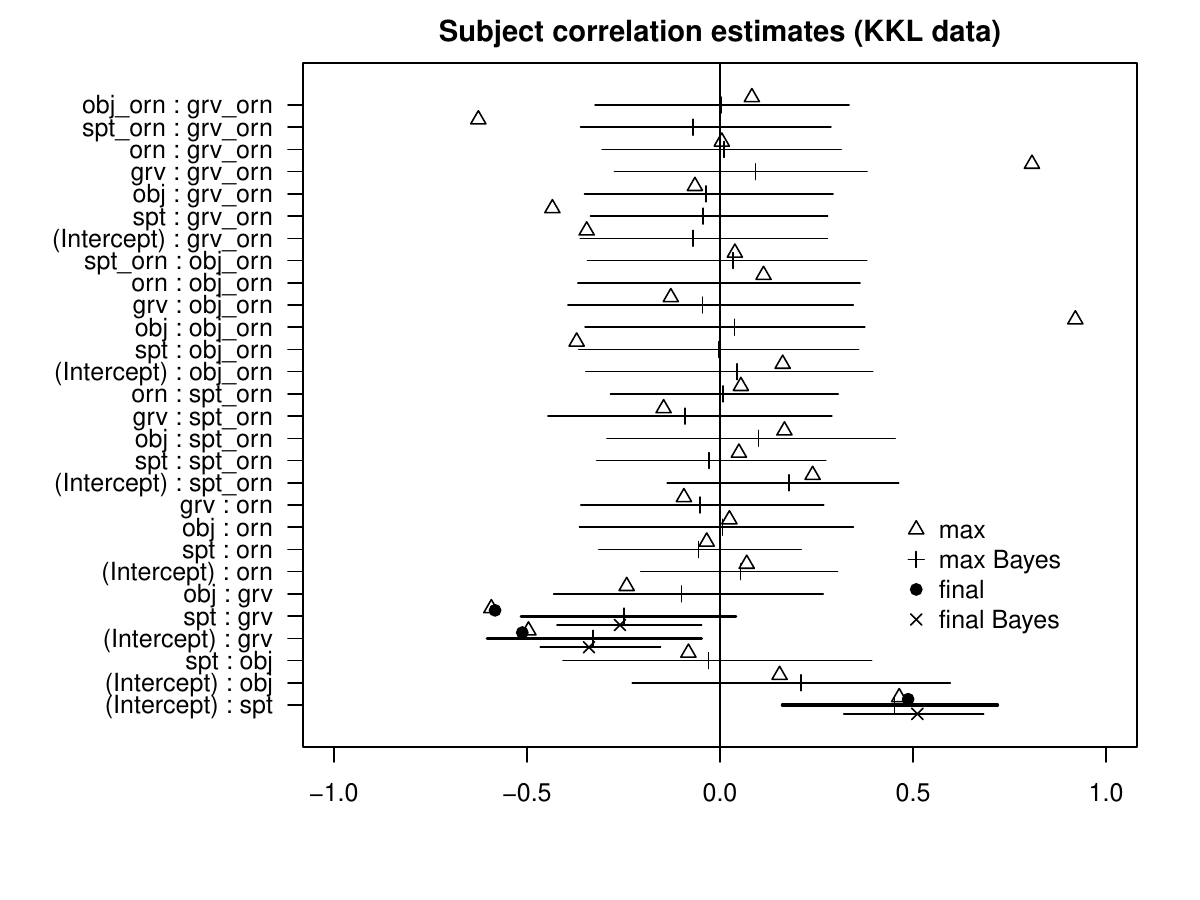} 

\end{knitrout}
\caption{\label{fig:varcompkkl} Top: The standard deviation estimates from the maximal and final models of the KKL data, and estimates and 95\% credible intervals of the maximal and final Bayesian hierarchical models. Bottom: Correlations of subject random effects in the maximal and final models of the KKL data, along with estimates and credible intervals from the maximal and final Bayesian model.}
\end{figure}

We start with the maximal linear mixed model including all
possible variance components and correlation parameters associated with the four within-subject contrasts in the random-effects structure. Note that there are no interactions between the three contrasts associated with the four levels of the cue-target relation factor. Also, as factor size was manipulated between subjects, this contrast does not appear in the random-effect structure. Thus, the random-effect structure comprises eight
variance components (i.e., the intercept estimating the grand mean of log reaction time, the three contrasts for the four types of cue-target relation, the contrast for the orientation factor, and three interactions) and $28$  correlation parameters ($8 \times 7 / 2$)--also a very complex model.  
The maximal model converges with a warning:

\begin{verbatim}
maxfun < 10 *length(par)^2 is not recommended
\end{verbatim}

\noindent 
This suggests that we may be asking too much of these data. Nevertheless, at a first glance, model parameters look
reasonable. As shown in Figure~\ref{fig:varcompkkl},
none of the eight variance components are estimated at zero and none of the 28 correlation parameters are at the boundary (i.e., none assume values of $+1$ or $-1$). The PCA, however, indicates that the maximal model is overparameterized: two dimensions contribute 0\% variance explained.

\begin{figure}
\centering
\begin{knitrout}
\definecolor{shadecolor}{rgb}{0.969, 0.969, 0.969}\color{fgcolor}
\includegraphics[width=\maxwidth]{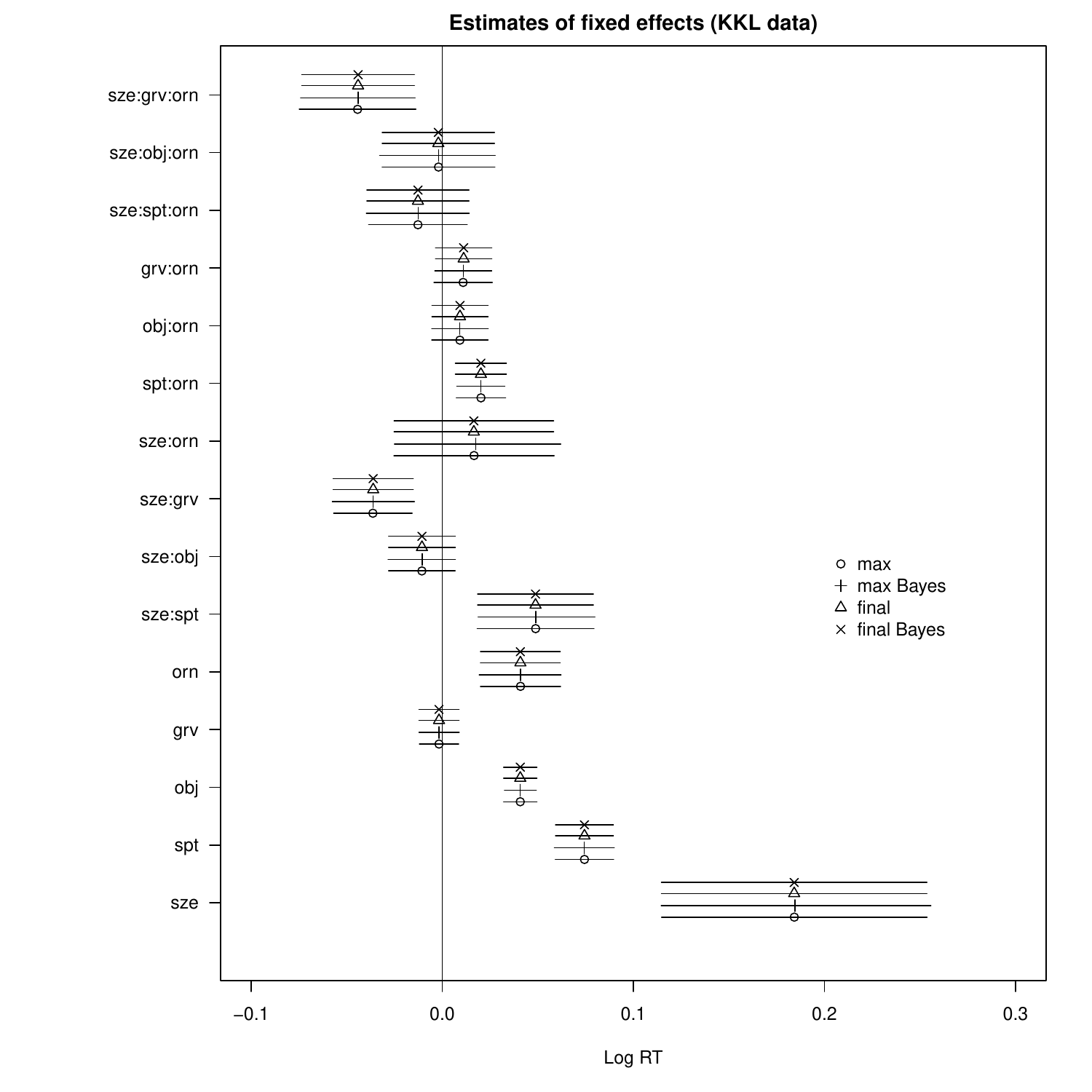} 

\end{knitrout}
\caption{\label{fig:fixefkkl} The estimates and 95\% confidence intervals for the fixed effects in the maximal and final models of the Kliegl et al.\ 2015 data. Also shown are estimates and 95\% credible intervals from maximal and final Bayesian hierarchical linear models.}
\end{figure}

\subsubsection{Zero-correlation-parameter linear mixed models}

The problem of overidentification persists in the zero-correlation parameter model (LMM). In the PCA, we still have only seven of eight non-zero components and one of them accounts for less that 1\% of the variance. Thus, the LMM still is too complex for the information contained in the data of this experiment.

\subsubsection{Dropping variance components to achieve model identification}

The estimates of variance components suggest that there is very little reliable variance associated with the interaction between object and orientation contrasts and for the interaction between gravitation and orientation. Dropping these two variance components from the model and refitting leads to an identified LMM. Thus, the data of this experiment support six variance components, in agreement with the initial PCA of the maximal model.

\subsubsection{Testing non-significant variance components}

Given an identified LMM, we test whether removal of any of the remaining variance components reduces the goodness of fit in an LRT. It turns out that all of these variance components are reliable. So we keep them in the model.

\subsubsection{Extending the reduced LMM with correlation parameters}

Having arrived at an identified reduced LMM, we expand the LMM and
check whether there are significant correlation parameters.\footnote{There is a slight risk that we removed a variance component that would have been significant with correlation parameters in the LMM, but we found no
evidence for this in additional analyses.} This model is also supported by the data: There is no evidence of degeneration. Moreover, the model fits significantly better than the zero-correlation parameter model; $\chi_{15}^{2}
= 50, p < .01$. Thus, we would consider this LMM as an acceptable model. The results are documented in Figures \ref{fig:fixefkkl} and
\ref{fig:varcompkkl}.

\subsubsection{Pruning low correlation parameters}

The significant increase in goodness of fit when going from the reduced zero-correlation parameter model to the extended LMM suggests that there is significant information associated with the ensemble of correlation parameters.  Nevertheless, the object and orientation effects and the interaction between spatial and orientation effects are only weakly correlated with the mean as well as with spatial and gravitation effects. So we remove
these correlation parameters from the model. There is no loss of goodness of fit associated with dropping most of the correlation parameters; $\chi_{12}^{2} = 8.4, p = .75$.

\subsubsection{Summary}

The data from this experiment were a follow-up to an experiment reported by 
\citet{KlieglWeiDambacherYanZhou2011}. The statistical inferences in that article, especially also with respect to correlation parameters, were based on a maximal LMM. A reanalysis along the strategy described here revealed an overparameterization, involving a negative correlation between spatial and attraction effect and a positive correlation between mean and spatial effect.
The reanalysis of those early data is also part of the \texttt{RePsychLing} package accompanying the present article. The theoretically important negative and positive correlation parameters were replicated with the present experiment
in the absence of problems with model complexity (see
\citealp{KlieglKuschelaLaubrock2014} for further discussion.)  

\subsubsection{An analysis of  Kliegl et al. (2015) using a Bayesian LMM}

We also fit a maximal Bayesian linear mixed model using \texttt{rstan}. As in the analysis of the Kronm\"uller and Barr (KB) data, this also showed that the same variance components whose credible intervals do not include zero  were the ones that the iterative procedure identified as suitable for inclusion. As in the KB data, notice that the means of the correlation estimates in the Bayesian model tend to be closer to zero than those from \texttt{lme4}. This is because the prior on the correlation matrix has most of its probability mass around zero. If the sample size is small, the prior will dominate in determining the posterior distribution of the correlations; but if there is sufficient data, and if a non-zero correlation is truly present, the mean of the posterior distribution could be different from zero. We see this in the case of three correlation parameters (Figure \ref{fig:varcompkkl}).  For comparison, we also show the estimates from a Bayesian model corresponding to the final LMM chosen in the `lme4' analysis presented above.

In sum, the Bayesian analysis independently validates the conclusions based on the PCA-based approach described above.

\section{Discussion}

An important goal in statistical analysis of empirical data is the avoidance of overfitting.  Any given data-set can tolerate only a limited number of parameters.  Mixed-effects modeling is no exception.  In the statistical literature on fitting mixed-effects modeling (see, e.g., \citealp{PB},\citealp{GaleckiBurzykowski2013},\citealp{BatesMaechlerBolkerWalker2015}), the
approach taken is one in which variance components are added to the model step by step, typically driven by theoretical considerations.   
The recommendation of \citet{BarrLevyScheepersTily13} to fit \texttt{maximal}
models with all possible random effect components included comes from a very different tradition in which statistics is used to provide a verdict on significance in factorial designs.  The authors based their recommendation on a simulation study indicating that anti-conservative results were best avoided by fitting models with as rich a random effects structure as possible.  

It is indeed important to make sure that the proper variance components are included in the mixed model.  Failure to do so may result in anti-conservative conclusions.  However, the advice to ``keep it maximal'' often creates hopelessly over-specified random effects because the number of correlation parameters to estimate rises quickly with the dimension of the random-effects vectors.  The information in the data may not be sufficient to support
estimations of such complex models and may result in singular covariance matrices, even when the LMM is identifiable in principle. In this case, we need to replace the complex LMM specification by a more parsimonious one.

With an iterative reduction of the complexity of a degenerate maximal model, one can obtain a model in which the estimated parameters are in line with the information present in the data. We proposed (1) to use PCA to determine the dimensionality of the variance-covariance matrix of the random-effect structure, (2) to initially constrain correlation parameters to zero, especially when an initial attempt to fit a maximal model does not converge,
and (3) to drop non-significant variance components and their associated correlation parameters from the model.  Each of these reductions may lead to a significant loss in goodness of fit according to LRTs for nested models, in which case this clarifies that the parameter is actually well-supported by information in the data.

Importantly, failure to converge is not due to defects of the estimation algorithm, but is a straightforward consequence of attempting to fit a model that is too complex to be properly supported by the data.  We have presented examples showing that the problem of overspecification may arise irrespective of whether estimation is based on maximum likelihood or on Bayesian hierarchical modeling.   Furthermore, even under convergence, overparameterization may lead to uninterpretable models, which is
why we developed a diagnostic tool for detecting overparameterization.

What one typically finds for overspecified, degenerate models --- degenerate because they afford no predictive power beyond an identified model for the same data --- is that the presence of superfluous variance components has minute effects on the estimates of the variability of the fixed-effects estimates. They may occasionally affect standard errors in the decimals, pushing a
$p$-value below or lifting one above some supposedly `critical' level of $\alpha$. This should not make a difference as far as conclusions regarding the fixed effects parameters are concerned \citep{BatesMaechlerBolkerWalker2015}.   In fact, comparing parsimonious models with the maximal models discussed by Barr et al.\ (see
the \texttt{RePsychLing} package for R for full details on the analyses),
there is not a single instance where conclusions about the fixed-effect predictors diverge.   Thus, for these real data sets, it is not necessary to aim for maximality when the interest is in a confirmatory analysis of factorial contrasts.

If for some reason it is critical to establish the reliability of a specific variance component or correlation parameter, the most promising approach, where feasible, is to collect more data. Beyond that we must mind the Sunset Salvo \citep{Tukey1986}: ``The combination of some data and an aching desire for an answer does not ensure that a reasonable answer can be extracted from a given
body of data'' (p.74--75). 

What then about the simulation studies on which Barr et al. base their recommendations for maximality?  Several issues arise here, which all relate to how representative these simulations are with respect to real data sets. First, the simulations implement a factorial contrast that is atypically large compared to what is found in natural data.  Second, and more importantly, the
correlations in the random effects structure range from $-0.8$ to $+0.8$.  Such large correlation parameters are indicative of overparameterization.  They hardly ever represent true correlations in the population.  As a consequence, these simulations do not provide a solid foundation for recommendations about how to fit mixed-effects models to empirical data.  

In summary, maximal models are not necessary to protect against
anti-conservative conclusions.  This protection is fully provided by comprehensive models that are guided by realistic expectations about the complexity that the data can support.  In statistics, as elsewhere in science, parsimony is a virtue, not a vice.  

 \bibliographystyle{rss}
 \bibliography{references}
 
\end{document}